\documentclass[pra,twocolumn,superscriptaddress,nofootinbib]{revtex4-2}
\usepackage{graphicx}
\usepackage{bm}
\usepackage{amssymb,amsmath,mathrsfs}
\usepackage{color}
\usepackage{amstext}
\usepackage[english]{babel}
\usepackage{latexsym}
\usepackage{array}   
\usepackage{multirow}         
\usepackage[usenames,dvipsnames]{xcolor}
\usepackage[colorlinks=true,citecolor=Blue,linkcolor=RubineRed,urlcolor=Blue]{hyperref}
\usepackage[all]{xy}     
\usepackage[usenames,dvipsnames]{xcolor}

\newcommand{\red}[1]{{\color{red} #1}}
\begin{document}

\title{Spin Glass Dynamics on Complex Hardware Topologies: A Bond-Correlated Percolation Approach}

\author{Viviana Gómez}
\email{v.gomez6@uniandes.edu.co}
\affiliation{Departamento de F{\'i}sica, Universidad de los Andes, A.A. 4976, Bogot\'a D.C, Colombia}

\author{Gabriel Téllez}
\email{gtellez@uniandes.edu.co}
\affiliation{Departamento de F{\'i}sica, Universidad de los Andes, A.A. 4976, Bogot\'a D.C, Colombia}

\author{Fernando J. Gómez-Ruiz}
\email{fegomezr@fis.uc3m.es}
\affiliation{Departamento de F\'isica, Universidad Carlos III de Madrid, Avda. de la Universidad 30, 28911 Legan\'es, Spain}

\begin{abstract}

Understanding how frustration and disorder shape relaxation in complex systems is a central problem in statistical physics and quantum annealing. Spin-glass models provide a natural framework to explore this connection, as their energy landscapes are governed by competing interactions and constrained topologies. We investigate the non-exponential relaxation behavior of spin glasses on network architectures relevant to quantum annealing hardware—such as finite size Chimera, Pegasus, and Zephyr graphs—where embedding constraints and finite connectivity strongly modulate the distribution of barriers and metastable states. This slow relaxation arises from the combined effects of frustration and disorder, which persist even beyond the conventional spin-glass transition. Within the Fortuin–Kasteleyn–Coniglio–Klein (FKCK) cluster formalism, the appearance of unfrustrated cluster regions gives rise to multiple relaxation scales, as distinct domains follow different dynamical pathways across a rugged energy landscape. This framework enables a more comprehensive characterization of spin-glass energy landscapes and offers valuable insight into how topological constraints and disorder jointly govern relaxation dynamics, providing quantitative benchmarks for evaluating the performance and limitations of quantum annealing architectures.

\end{abstract}


\maketitle

\section{Introduction}
In the realms of quantum computing and quantum simulation, a fundamental yet highly nontrivial problem is the preparation of ground states of interacting Hamiltonians~\cite{NORI_RMP14,Ekert_RMP96, Kishor_RMP22, Sam_RMP20}. Quantum annealing (QA) has established itself as a theoretical and practical paradigm for addressing computationally demanding combinatorial optimization problems, with experimental implementations based on networks of coupled qubits that realize programmable interactions~\cite{Hidetoshi_QA_PRE98,Aharonov04,Lidar_RMP18}. Several hardware architectures have been developed following this principle, featuring connectivity patterns such as the D-Wave Chimera, Pegasus and Zephyr graphs, which represent successive efforts to increase qubit connectivity, embedding efficiency, and scalability~\cite{boothby2019pegasus,boothby2021zephyr,pelofske2025comparing}. These programmable quantum processors, exemplified by D-Wave's annealing platforms~\cite{King_Science25,King2023-pc,King2022-os, Bando_PRR20}, have enabled the exploration of non-equilibrium spin-glass dynamics, where phenomena akin to the Kibble–Zurek mechanism govern defect formation and relaxation processes during the annealing trajectory~\cite{Zurek96a,*Zurek96b,*Zurek96c, Kibble76a,*Kibble76b, DKZ13,GomezRuiz20}.

Spin glasses, disordered magnetic systems characterized by competing ferromagnetic and antiferromagnetic interactions, offer a natural framework for formulating hard combinatorial optimization problems~\cite{santoro2002theory}. Their Hamiltonians, typically expressed as Ising models with quenched disorder in the couplings, capture rugged energy landscapes with a multitude of nearly degenerate local minima separated by large barriers~\cite{Chowdhury,Knysh2016,Andras_PRR25}. In the context of quantum adiabatic computing, such Hamiltonians serve as benchmark instances, as the adiabatic evolution seeks the ground state configuration within these complex landscapes~\cite{Bernaschi2024}. Owing to their intrinsic complexity, spin-glass models pose a formidable challenge for classical algorithms and, at the same time, provide an ideal testbed for assessing the potential advantages of quantum approaches~\cite{heim2015quantum}.

Spin glasses exhibit a transition to a spin-glass phase at a characteristic temperature $T_{\rm SG}$, below which the system freezes into a disordered but stable configuration with no conventional long-range magnetic order. This transition marks the point where the spins become trapped in a complex landscape of metastable states, leading to a dramatic slowing down of dynamics and to history-dependent phenomena such as aging~\cite{sherrington1975solvable, gunnarsson1988dynamics, singh1986critical}. However, in glassy and frustrated systems, it is often observed that the dynamical behavior changes well before any static transition takes place~\cite{randeria1985low, franzese1999precursor}. Specifically, spin–spin autocorrelations, which decay exponentially at high temperatures, develop a stretched or nonexponential form below a crossover scale $T^*$~\cite{ogielski1985dynamics, pezzella1997spin}. This dynamical crossover can arise from two main mechanisms: frustration and disorder. Percolation processes, such as the formation of Fortuin–Kasteleyn–Coniglio–Klein (FKCK) clusters, provide a microscopic explanation for the increasing complexity of the accessible phase space in the presence of frustration. Below $T_p$, the temperature at which these clusters percolate, the dynamics slow down and relaxation becomes heterogeneous, even in fully frustrated models without explicit disorder~\cite{franzese1998phase}. In the presence of disorder, rare regions can locally order and give rise to a Griffiths phase, with a broad distribution of relaxation times. In both scenarios, the net effect is that the autocorrelation function deviates from a simple exponential, reflecting the coexistence of multiple timescales and the onset of slow dynamics. Determining whether $T^*$ coincides with the ferromagnetic Ising critical temperature $T_C$ or with the percolation temperature $T_p$ remains a challenging numerical problem, as the two scales typically lie in close proximity.

Analogous to what occurs in glass-forming liquids, the emergence of non-exponential relaxation in spin glasses is closely related to changes in how the system explores its potential energy landscape~\cite{glotzer2000potential, sastry1998signatures, jonsson1988icosahedral}. At high temperatures, the average energy decreases only weakly with decreasing $T$, indicating that the system explores shallow regions of the landscape. As the spin-glass transition temperature $T_{\rm SG}$ is approached, however, the energy begins to decrease more rapidly, reflecting that the system becomes confined to progressively deeper minima. This regime of enhanced energy relaxation coincides with the temperature range where the spin autocorrelation function transitions from exponential to non-exponential decay, signaling the onset of glassy dynamics.

In experimental realizations of these systems, such as those implemented on quantum annealing hardware, the situation becomes even more intricate due to the unavoidable presence of finite operating temperatures. Practical implementations of quantum annealers necessarily operate at finite temperatures, which imposes intrinsic limitations on their performance as scalable optimizers~\cite{Albash_PRL17,Abbas2024}. Recent studies have shown that unless the operating temperature decreases appropriately with system size, the device cannot reliably sample ground states or exploit quantum tunneling effectively~\cite{Crosson2021,MehtaPRA25,Lidar_PRL25}. Consequently, the competition between quantum and thermal fluctuations becomes a key factor shaping the relaxation dynamics of spin-glass systems realized on current annealing processors.

The central idea of QA is to gradually quench quantum fluctuations in order to minimize a target Hamiltonian that encodes a complex optimization problem~\cite{rajak2023quantum}. In principle, the quantum wave function may tunnel through barriers in the free-energy landscape, suggesting a possible advantage over classical methods such as simulated annealing (SA), where thermal fluctuations are progressively reduced to reach low-energy states~\cite{kirkpatrick1983optimization}. However, previous studies have suggested that, for spin-glass systems defined on the Chimera topology, the structure of the energy landscape may actually favor thermal strategies like SA, since the spin-glass phase is stable only at zero temperature~\cite{katzgraber2015seeking, katzgraber2014glassy}. In this work, we revisit the interplay between topology, frustration, and relaxation in spin-glass models defined on D-Wave–inspired architectures. By analyzing the emergence of non-exponential relaxation through classical Monte Carlo simulations and interpreting the results within the framework of percolation theory, we aim to clarify how network connectivity shapes dynamical slowing down. Our objective is not to demonstrate a quantum speedup—a question that remains open—but to deepen the understanding of frustration in complex networks and provide insight that may guide future developments in both classical and quantum optimization.

The structure of this article is as follows: In Sec.~\ref{sec:percolation_transition}, we begin by introducing the fundamental concepts of percolation of FKCK clusters in spin glasses and their relation to the phase space of glassy systems. In Sec.~\ref{sec:methodology}, we describe the methodology employed to obtain the results. In Sec.~\ref{results}, we present an analysis of the percolation properties of these clusters in the Chimera, Pegasus, and Zephyr topologies, together with a study of the associated spin–spin autocorrelations. Finally, we discuss the implications and relevance of these results for understanding glassy dynamics in quasi-2D networks and for assessing their role as benchmarks in the search for quantum advantage.

\section{Percolation transition}\label{sec:percolation_transition}

In the study of critical phenomena, the FKCK cluster formalism provides a geometric framework that links thermodynamic correlations to percolation properties~\cite{fortuin1972random}. In the ferromagnetic Ising model, this approach introduces bonds between parallel nearest-neighbor spins of coupling strength $J$, which are connected with a temperature $T$ dependent probability
\begin{equation}
p(T) = 1 - \exp\left[-\frac{2J}{k_B T}\right].
\label{eq:probability}
\end{equation}
By summing over spin configurations weighted by the Boltzmann factor, the partition function can be reformulated as a sum over bond configurations. Within this representation, FKCK clusters describe correlated spin fluctuations that percolate precisely at the Ising critical point, exhibiting the correct critical exponents. This geometric framework has proven particularly powerful for analyzing how network topology shapes the onset and universality of critical behavior~\cite{gomez2024self}.


When extending this approach to disordered and frustrated models such as spin glasses, important modifications appear. Consider the Ising spin glass Hamiltonian
\begin{equation}
H \;=\; -J \sum_{\langle ij\rangle} \varepsilon_{ij} S_i S_j ,
\end{equation}
where the quenched random couplings \(\varepsilon_{ij} = \pm 1\) represent ferromagnetic or antiferromagnetic interactions. Following the FKCK formalism, a bond is placed between neighboring spins that satisfy their local interaction ($\varepsilon_{ij} S_i S_j = +1$) with probability $p$. However, the essential difference from the ferromagnetic case arises from frustration: along a closed loop for which $\prod_{\langle ij \rangle \in \text{loop}} \varepsilon_{ij} = -1$, no spin configuration can simultaneously satisfy all interactions. Such loops are said to be frustrated, and bonds cannot be assigned in a way that closes them.

As a result, the spin glass partition function can be expressed as
\begin{equation}
Z \;=\; \sum_{C^*} e^{\mu\,b(C)}\, 2^{N(C)},
\end{equation}
where $\mu = \log[p/(1-p)]$, $b(C)$ is the number of bonds in the cluster configuration $C$ and $N(C)$ is the total number of clusters in that configuration. The sum $\sum_{C^*}$ runs only over bond configurations that do not contain frustrated loops. At zero temperature, the ground state corresponds to the maximal number of bonds consistent with the no–no-frustration constraint, turning the problem into a geometrical optimization under frustration~\cite{coniglio1991cluster, de1999cluster, franzese1998phase}.

This distinction has significant consequences. In the ferromagnetic case, clusters directly mirror thermodynamic fluctuations, whereas in spin glasses correlations are mixed in sign: two spins may be positively correlated if the path connecting them involves an even number of negative bonds, or negatively correlated if the path involves an odd number. These competing contributions partially cancel, meaning that clusters defined via connectivity no longer represent thermodynamic correlations. Nevertheless, connectivity is always non-negative, so clusters defined by bonds that do not close a frustrated loop can still percolate. This occurs at a temperature $T_p$ higher than the actual spin glass critical temperature $T_{\rm SG}$~\cite{munster2024spin, munster2023cluster}. This percolation threshold $T_p$ can be mapped to the thermodynamic transition of a ferromagnetic $2s$-state Potts model  (see Refs.~\cite{franzese1998phase, coniglio1991cluster, franzese1999precursor}). This exact equivalence between Potts thermodynamics and the percolation of correlated clusters, allows one to interpret $T_p$ as a Potts-like transition. In the three–dimensional binary Ising spin glass, numerical simulations show $T_p \approx 3.95 J/k_B$~\cite{de1991percolation} while $T_{\rm SG} \approx 1.11 J/k_B$~\cite{katzgraber2006universality}.

\begin{figure}[h!]
    \begin{center}
    \includegraphics[width=\linewidth]
    {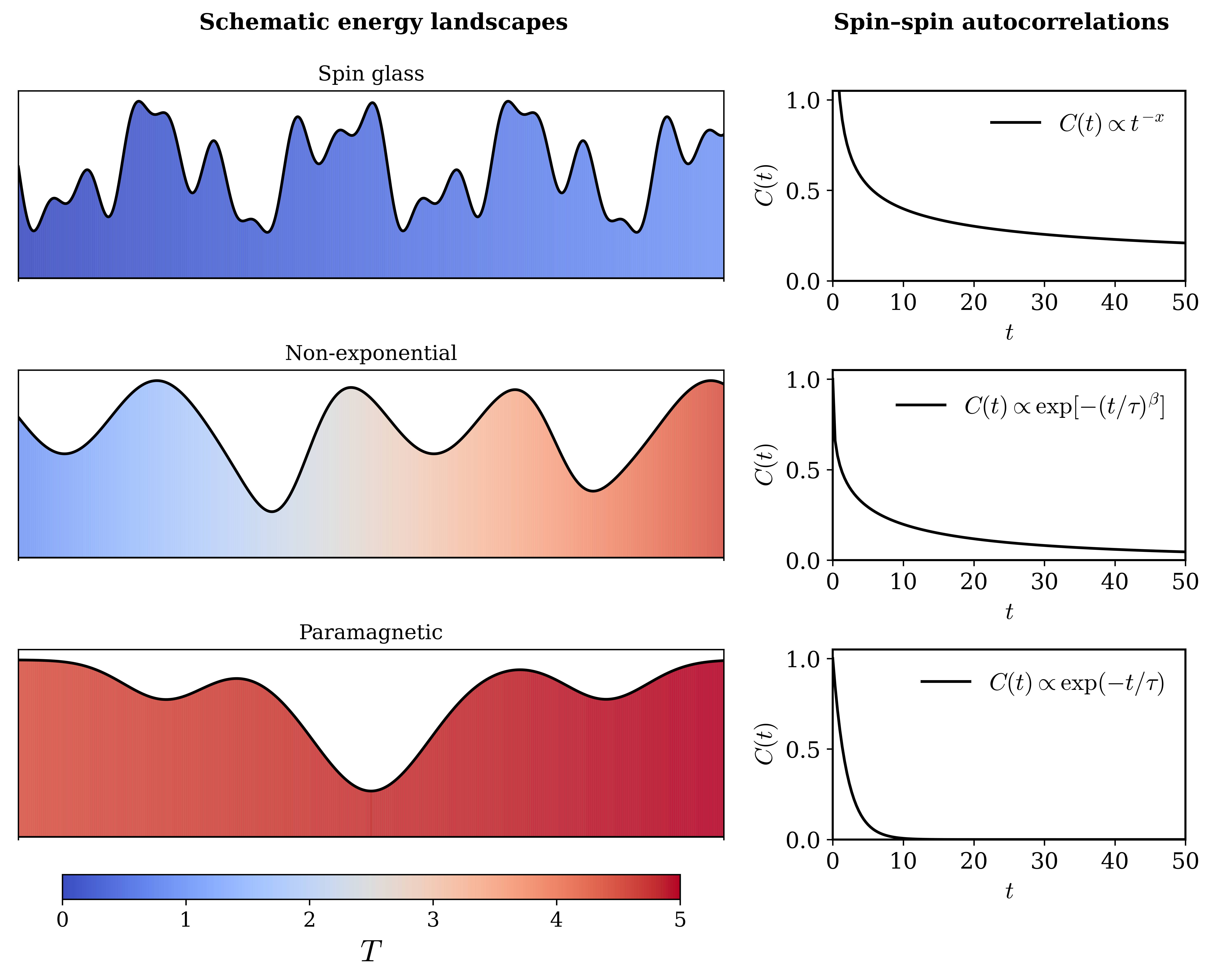}
    \end{center}
    \caption{Illustrative energy landscapes (left) and their corresponding spin–spin autocorrelations (right). Below the spin glass transition, the energy landscape becomes extremely rugged, with many deep valleys separated by high barriers; ergodicity is broken and spin autocorrelations decay slowly. Below the Potts (percolation) transition, frustration persists, while below the Griffiths temperature, disorder induces rare correlated regions—both mechanisms give rise to stretched-exponential relaxation. Above these phases, the system is paramagnetic, exhibiting exponential decay and a simple energy landscape.}
    \label{fig:fig1}
\end{figure}

The phase diagram of a spin glass can be interpreted as follows. Below the critical temperature $T_{\rm SG}$, the system enters the spin-glass phase, where ergodicity is broken and spin autocorrelations decay slowly, often following a power-law form. Above $T_{\rm SG}$, however, the system already exhibits non-exponential relaxation, which can arise from two distinct mechanisms. For temperatures below $T_p$, frustration generates a corrugated energy landscape that induces stretched or non-exponential relaxation, even in the absence of disorder (see Fig.~\ref{fig:fig1}). In contrast, below $T_c$---which coincides with the critical temperature of an Ising ferromagnet---quenched disorder becomes relevant, giving rise to rare, strongly correlated regions that define the Griffiths phase. Both the Potts transition at $T_p$ and the Griffiths transition at $T_c$ contribute to the emergence of non-exponential relaxation. Since typically $T_c > T_p$, the disorder-driven Griffiths transition dominates and tends to mask the effects of frustration. In the limiting case of a fully frustrated model without disorder, only the Potts transition at $T_p$ remains, clearly signaling the onset of non-exponential relaxation~\cite{franzese1999precursor}.

\section{Methodology}\label{sec:methodology}

In this study, we investigate the percolation properties and dynamical behavior of spin glass models defined on network topologies relevant for QA: Chimera, Pegasus, and Zephyr graphs, as well as a three–dimensional cubic lattice for comparison. Specifically, we determine the percolation threshold $T_p$ for different system sizes in each topology. To model disorder and frustration, we employ the Edwards–Anderson (EA) spin glass with Gaussian couplings. In this formulation, the Hamiltonian is given by
\begin{equation}
H = - \sum_{\langle i,j \rangle} J_{ij} S_i S_j,
\end{equation}
where $S_i = \pm 1$ are binary Ising spin variables to be optimized, $J_{ij}$ are quenched random coupling strengths drawn from a Gaussian distribution with zero mean and unit variance, and $\langle i,j \rangle$ denotes the set of connected pairs of spins in the underlying graph of the model.

\begin{figure*}[t!]
\includegraphics[width=1.0\linewidth]{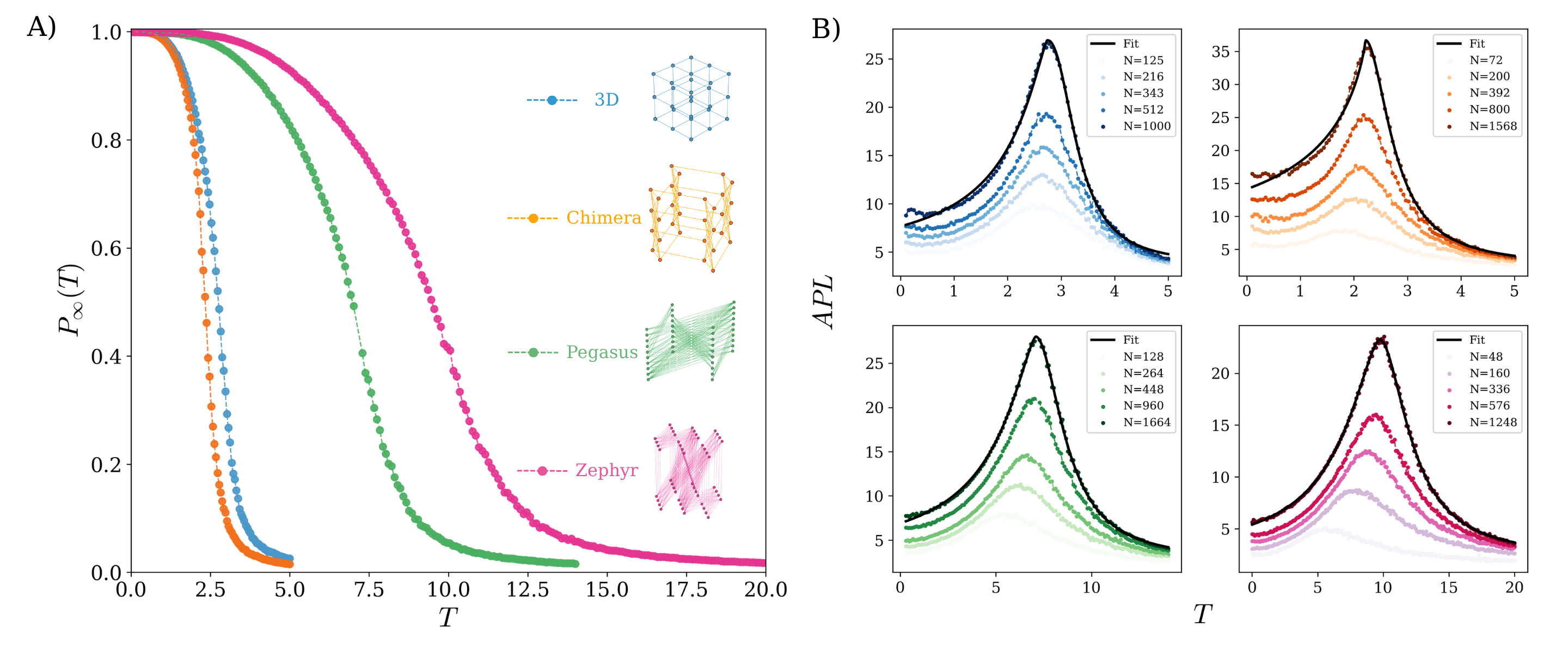}
\caption{{\textbf{A)} Spanning probability $P_{\infty}(T)$ for the largest system size of each lattice: Blue – 3D ($N=1000$), Orange – Chimera ($N=1568$), Green – Pegasus ($N=1664$), Pink – Zephyr ($N=1248$). 
   \textbf{B)} Average path length for all system sizes studied in each network. The results are well described by a finite-size scaling analysis, which allows us to estimate the percolation temperature $T_p$ in the thermodynamic limit (see Table~\ref{tab:combined_parameters}). The distances are weighted by the coupling strengths $J_{ij}$, but this weighting does not affect the position of the APL peak, only its absolute value. For the largest system sizes, the APL is fitted according to Eq.~\eqref{eq:APL_model}}.
    \label{fig:fig2}}
\end{figure*}

To determine the percolation threshold $T_p$, we compute the spanning probability $P_{\infty}(p)$, defined as the fraction of spins belonging to the largest connected cluster over the total number of spins in the system. Clusters are formed by connecting pairs of spins according to the bond–activation rule given in Eq.~\eqref{eq:probability}, with the difference that in the spin glass case the value of each coupling $J_{ij}$ is explicitly taken into account when deciding whether a bond is activated.
Near the percolation threshold $p_c$, the spanning probability follows the finite–size scaling relation
\begin{equation}
P_{\infty}(p) \simeq \tilde{P}_{\infty}\!\left[ N^{1/\nu}(p - p_c) \right],
\end{equation}
where $N$ denotes the system size and $\nu$ is the critical exponent of the correlation length.  According to Eq.~\eqref{eq:probability}, this probability can be viewed as a function $P(T)$; for clarity, from now on we will work with the variable $T$ under this transformation. In practice, the transition temperature $T_p$ is estimated from the crossing point of the curves $P_{\infty}(T)$ obtained for different system sizes $N$ ~\cite{coniglio1991cluster}.\\

In addition, we evaluate the average path length (APL) as a measure of the network's effective connectivity under spin interactions. For a system of $N$ spins, the APL is defined as  
\begin{equation}
\langle \ell \rangle = \frac{2}{N(N-1)}
\biggl[
\Bigl\langle 
\sum_{i<j} \ell_{ij}
\Bigr\rangle_{T}
\biggr]_{J},
\end{equation}
where $\ell_{ij}$ denotes the shortest path between spins $i$ and $j$. The brackets $\langle \cdots \rangle_T$ and $[\cdots]_J$ represent thermal and disorder averages, respectively, taken over independent realizations of temperature histories and quenched couplings. In disconnected graphs, only connected components contribute to the sum, and if the network is completely disconnected, the APL is defined to be zero. The paths are constructed according to the bond–activation probability defined in Eq.~\eqref{eq:probability}, ensuring that stronger couplings are more likely to contribute to the connectivity. The metric is weighted using $1/J_{ij}$ as the distance between connected spins, so that stronger bonds correspond to shorter effective paths. When the network begins to fragment, the APL exhibits a pronounced peak, reflecting the disruption of paths between distant spins.

After identifying the percolation temperature $T_p$, we compute the spin–spin autocorrelation function for each network. This observable is defined as
\begin{equation}
    C(t) = \frac{1}{N} \sum_{i=1}^N \biggl[ \, \langle S_i(0) S_i(t) \rangle_T \, \biggr]_{J},
    \label{eq:correlation}
\end{equation}
where $S_i(0)$ denotes the initial configuration, and the average is taken over both thermal histories and quenched disorder realizations drawn from the Gaussian EA distribution. The correlations were measured using a classical single–spin Monte Carlo flip dynamics after ensuring that the system had reached equilibrium. All temperatures are reported in units of $\sigma_J/k_B$, where $\sigma_J$ is the standard deviation of the coupling distribution.
\section{Results}\label{results}
\subsection{Spin glass percolation transition}

\begin{table*}[t]
\centering
\setlength{\tabcolsep}{24pt} 
\caption{{\bf Summary of characteristic parameters for each topology}. $T_{\rm SG}$ denotes the spin-glass transition temperature, $T_p$ the Potts (percolation) transition temperature, and $T_C$ the Griffiths (ferromagnetic) transition temperature. $\gamma_-$ and $\gamma_+$ correspond to the asymmetric peak exponents for the largest simulated system size $N$. \label{tab:combined_parameters}}
\vspace{6pt}
\begin{tabular}{lcccccc}
\hline\hline
\textbf{Graph} & $N$ & $T_{SG}$ & $T_p$ & $T_C$ & $\gamma_-$ & $\gamma_+$ \\
\hline
3D       & 1000 & 0.95$^{a}$ & 2.76$^{\red{*}}$ & 4.51$^{b}$ & 1.26$^{\red{*}}$ & 2.23$^{\red{*}}$ \\
Chimera  & 1568 & 0.00$^{c}$ & 2.25$^{\red{*}}$ & 4.16$^{d}$ & 0.84$^{\red{*}}$ & 1.89$^{\red{*}}$ \\
Pegasus  & 1664 & 0.00$^{e}$ & 7.29$^{\red{*}}$ & 12.60$^{f}$ & 1.41$^{\red{*}}$ & 1.93$^{\red{*}}$ \\
Zephyr   & 1248 & 0.00$^{g}$ & 10.02$^{\red{*}}$ & 16.23$^{\red{*}}$ & 1.58$^{\red{*}}$ & 1.86$^{\red{*}}$ \\
\hline\hline
\end{tabular}
\vspace{7pt}
\noindent\begin{minipage}{\textwidth}
\small
$^{a}$Ref.~\cite{katzgraber2006universality} \hspace{1em}
$^{b}$Ref.~\cite{ferrenberg2018pushing} \hspace{1em}
$^{c}$Refs.~\cite{boothby2019pegasus,katzgraber2014glassy,jauma2024exploring} \hspace{1em}
$^{d}$Ref.~\cite{katzgraber2014glassy} \hspace{1em}
$^{e}$Refs.~\cite{boothby2019pegasus,jauma2024exploring} \hspace{1em}
$^{f}$Ref.~\cite{boothby2019pegasus} \hspace{1em}
$^{g}$Ref.~\cite{jauma2024exploring} \hspace{1em} $^{\red{*}}$ This study
\end{minipage}
\end{table*}
Figure~\ref{fig:fig2} shows the spanning probability $P_{\infty}(T)$ and the average path length for each of the four network topologies. The corresponding percolation temperatures, obtained using finite-size scaling (FSS) of the APL, are reported in Table~\ref{tab:combined_parameters}, together with a comparison to the other characteristic phase transition temperatures ($T_{\rm SG}$ and $T_C$). The APL can be modeled by an asymmetric peak function of the form
\begin{align}\label{eq:APL_model}
f(T) &= \frac{A}{1 + \left| \frac{T-T_p}{w_{\pm}} \right|^{\gamma_\pm}} + B, \\
\text{with} \quad w_{\pm}, \gamma_{\pm} &=
\begin{cases}
w_+, \gamma_+ & \text{if } T > T_p,\\
w_-, \gamma_- & \text{if } T < T_p. 
\end{cases}\notag
\end{align}
where $T_p$ is the temperature corresponding to the peak of the APL. The values of the exponents $\gamma_\pm$ for the largest system sizes are reported in Table~\ref{tab:combined_parameters}.

For the Zephyr network, the critical temperature $T_C$ was additionally estimated using the same FSS analysis applied to the APL, but considering the ferromagnetic case, i.e., with all couplings set to $J_{ij}=1$. Due to the relatively small network sizes accessible in our simulations, the reported values may slightly underestimate their true thermodynamic limits. Nonetheless, the extracted trends provide a reliable reference for comparing the effective behavior across architectures. 

As expected, the percolation temperature $T_p$ and the Griffiths transition temperature $T_C$ increase with the overall connectivity of the network. Among the studied topologies, Chimera exhibits the lowest $T_p$, consistent with its relatively low average degree and modular connectivity pattern. Interestingly, Chimera also displays the highest APL values immediately before the fragmentation of the percolating cluster, indicating that information or correlations must traverse longer paths to maintain global connectivity. This behavior reflects the presence of sparse inter-cell couplings, which create local bottlenecks and enhance the effective distance between distant regions of the network. In contrast, Zephyr, with its denser and more uniform inter-qubit connections, shows the largest $T_p$ and $T_C$, consistent with a more robust and strongly interconnected structure that delays fragmentation. 

\subsection{Relaxation analysis}\label{subsec:relaxation}

\begin{figure*}[t!]
    \begin{center}
    \includegraphics[width=\linewidth]
    {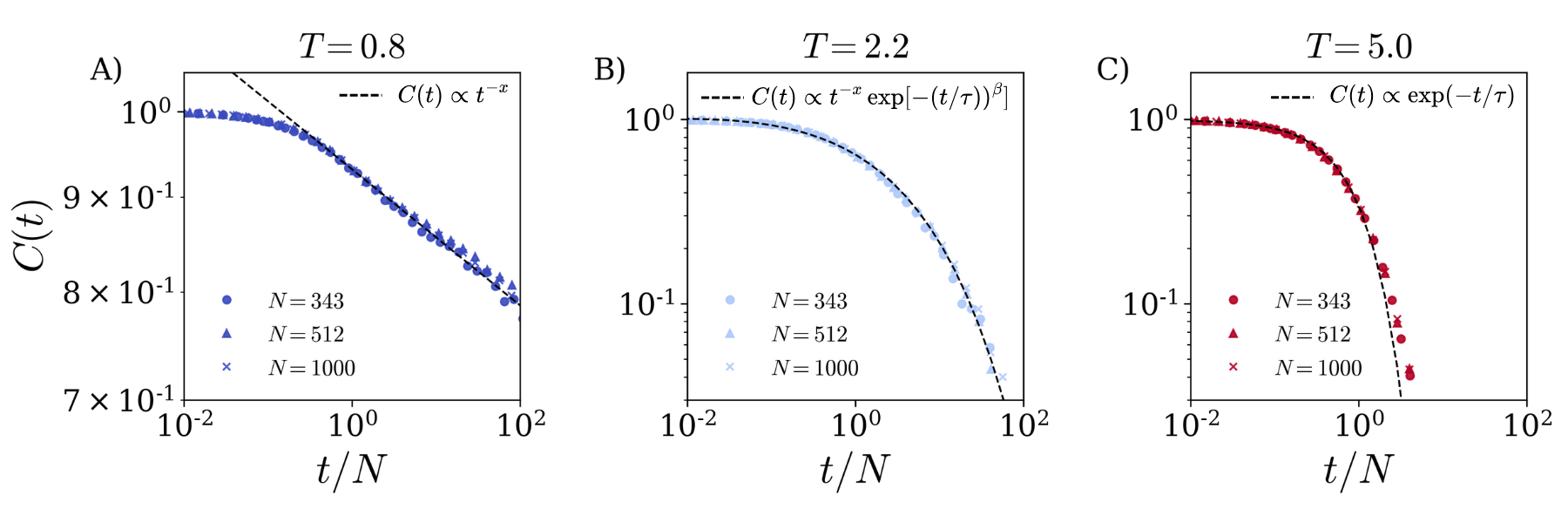}
    \end{center}
    \caption{Log–log plots of the spin–spin autocorrelation function (Eq.~\eqref{eq:correlation}) for different system sizes on a 3D network. \textbf{A)} $T=0.8$, spin-glass phase. \textbf{B)} $T=2.2$, non-exponential relaxation regime. \textbf{C)} $T=5.0$, paramagnetic phase.\label{fig:fig3}}
\end{figure*}

The relaxation behavior is characterized by examining the decay of the spin–spin autocorrelation function defined in Eq.~\eqref{eq:correlation} for each network topology. Simulations were performed using classical single–spin flip Metropolis Monte Carlo dynamics with random sequential updates. For each temperature, $N_\mathrm{conf}=50$ independent disorder realizations were done. Each simulation consisted of $10^6$ Monte Carlo steps per spin (MCS), with observables measured every MCS.

Relaxation in spin–glass systems has been extensively investigated. Analytical work by Randeria \textit{et al.}~\cite{randeria1985low} demonstrated that, even above the spin–glass transition, the decay of correlations under local spin–flip dynamics cannot be purely exponential, revealing an intrinsic constraint that reflects the presence of large, nearly unfrustrated spin clusters dominating the long–time dynamics. Complementary numerical studies by Ogielski~\cite{ogielski1985dynamics} showed that this slow relaxation is well captured by empirical, temperature–dependent forms of the type
\begin{equation}
C(t) = c \, t^{-x} \exp(-\omega t^{\beta}),
\label{eq:ogielski}
\end{equation}
where all parameters $c$, $x$, $\omega$, and $\beta$ vary with temperature. 

Figure~\ref{fig:fig3} shows the relaxation of the spin–spin autocorrelation function for a 3D network at different system sizes. The simulation time was measured in units of single–spin flips ($\delta t=1$) rather than full Monte Carlo sweeps, so that one unit of time corresponds to the attempted update of a single spin instead of all $N$ spins. This definition provides a much finer temporal resolution and allows us to capture early–time relaxation processes, which is particularly useful for highly connected networks where correlations evolve rapidly. To compare different system sizes, the horizontal axis was rescaled as $t/N$ since each Monte Carlo sweep would correspond, on average, to $N$ single–spin updates. With this rescaling, the relaxation curves for different $N$ collapse onto a single master curve.

At very short times, only a few spins have flipped and the system remains close to its initial configuration; correlations then decay slowly because relaxation is limited to local rearrangements that do not yet involve collective modes. The physically relevant behavior emerges in the long–time tails of the autocorrelation, where collective rearrangements driven by frustration and disorder dominate. For $T < T_{\rm SG}$ the relaxation follows a power–law behavior, for $T \lesssim T_C$ it is well described by Eq.~\eqref{eq:ogielski}, and for $T > T_C$ it reduces to a simple exponential decay.

\begin{figure*}[t!]
\includegraphics[width=\textwidth]{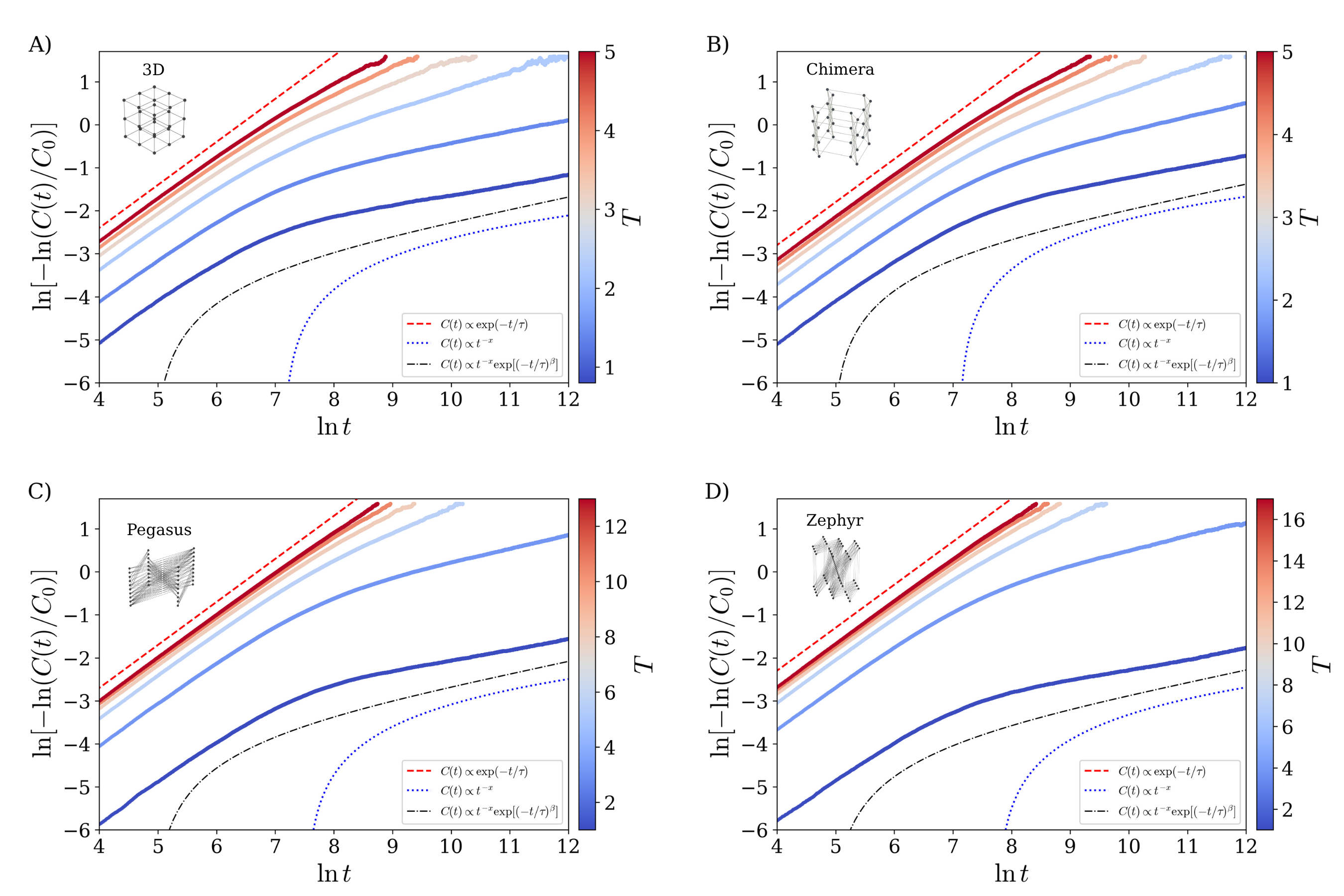}
\caption{\label{fig:fig4} Spin–spin autocorrelations (Eq.~\eqref{eq:correlation}) for different network topologies: \textbf{A)} 3D lattice with $N = 1000$ and $T_p = 2.84$, \textbf{B)} Chimera network with $N = 1568$ and $T_p = 2.27$, \textbf{C)} Pegasus network with $N = 1664$ and $T_p = 7.18$, and \textbf{D} Zephyr network with $N = 1248$ and $T_p = 9.78$. Here, $t$ corresponds to a single-spin Monte Carlo flip attempt rather than a full-lattice update, in order to obtain higher temporal resolution. The reference power-law curves were obtained from fits to the lowest-temperature data, yielding exponents $x_{\text{3D}} = 0.067$, $x_{\text{Chimera}} = 0.104$, $x_{\text{Pegasus}} = 0.050$, and $x_{\text{Zephyr}} = 0.040$.}
\end{figure*}

Given the relaxation behavior identified in Fig.~\ref{fig:fig3}, we extended the analysis to all network topologies using the largest available system sizes to test whether similar scaling trends persist across architectures. Figure~\ref{fig:fig4} displays the relaxation dynamics at different temperatures for all of the graphs. The autocorrelation function (Eq.~\eqref{eq:correlation}) is rescaled in the form $\ln[-\ln(C(t)/C(0))$, so that if $C(t) \propto \exp[-(t/\tau)^\beta]$, the data should collapse onto a straight line with slope $\beta$. As the temperature increases, the slope approaches $\beta = 1$, corresponding to a simple exponential decay. In intermediate regimes, the relaxation is better described by a stretched–exponential form with an algebraic prefactor as in Eq.~\eqref{eq:ogielski}, consistent with the behavior reported by Ogielski~\cite{ogielski1985dynamics}.

Accurately extracting the exponents $\beta$ and $x$ from this form is challenging because the fitting function involves four free parameters: $x$, $\beta$, $\tau$, and the prefactor amplitude $c$. These parameters are not independent, and significant correlations arise between them during the optimization process. In particular, variations in $\beta$ can be partially compensated by changes in $c$.

Notably, above the percolation temperature $T_p$, the behavior observed in Figure \ref{fig:fig4} gradually approaches a straight line, highlighting a progressive change in the spin–spin autocorrelations from a power-law decay to a simple exponential form. While this does not correspond to an abrupt transition, it clearly signals a smooth crossover in the relaxation dynamics. As the temperature increases beyond $T_p$, the influence of frustrated regions diminishes, and the dynamics become dominated by more uniform, unfrustrated spin domains. By the time the system approaches the critical temperature $T_C$, the curves nearly collapse onto each other, exhibiting almost identical behavior. This progressive crossover, rather than a sharp transition, is consistent with theoretical expectations: the effects of frustration and quenched disorder gradually lose relevance, and the system enters a regime characterized by simple paramagnetic relaxation, well captured by a single exponential decay. This trend is consistently observed across all the networks studied.

In Chimera, Pegasus, and Zephyr graphs, although no true spin-glass transition is reported in the thermodynamic limit, the low-temperature dynamics exhibit spin–spin autocorrelations remarkably similar to those observed in conventional spin-glass systems. For example, the darkest blue line in panel \textbf{A} of Figure \ref{fig:fig3}, corresponding to the 3D case, is located at $T = 0.8$, where this graph is still in a spin-glass phase (see Table~\ref{tab:combined_parameters}). In Ref.~\cite{jauma2024exploring}, Jauma \textit{et al.} showed that, although Chimera, Pegasus, and Zephyr graphs do not undergo a true spin-glass transition in the thermodynamic limit, a pseudo-critical temperature can still be defined for finite systems. This pseudo-critical temperature is obtained from the Binder cumulant, which gradually rises from zero in the paramagnetic regime to a finite value in the spin-glass regime, with the maximum of its derivative indicating the pseudo-critical temperature. However, it does not scale consistently with system size. For the same sizes we studied here, they reported $T_{\rm SC}^{\rm Chimera} = 0.904$, $T_{\rm SC}^{\rm Pegasus} = 2.130$, and $T_{\rm SC}^{\rm Zephyr} = 2.661$.

For the EA spin glasses studied in this work, the difference between the disorder-driven crossover temperature $T_C$ and the percolation temperature $T_p$ is significantly larger than in the case of a binary ($\pm J$) spin glass with equal probability couplings. For example, in a binary 3D graph, $T_p \approx 3.95$.  This is expected because the Gaussian distribution of couplings introduces a broader range of interaction strengths, increasing the effective disorder in the system. Consequently, energy barriers and local minima in the landscape are more heterogeneous, which delays the onset of uniform relaxation and shifts $T_C$ to higher values relative to $T_p$.

\section{Discussion}

The results show that the topology of the interaction network strongly influences the energy landscape of spin-glass systems. Different graphs—3D, Chimera, Pegasus, and Zephyr—display distinct critical temperatures ($T_{SG}$, $T_p$, $T_C$), highlighting how connectivity controls cluster formation, the spatial distribution of frustration, and the overall ruggedness of the landscape. More highly connected networks generally exhibit higher $T_p$ and $T_C$. Below $T_C$, the dynamics are dominated by slow relaxation due to the combined effects of frustration, nearly unfrustrated regions, and disorder, whereas above $T_C$ these effects are largely suppressed, leading to more uniform relaxation. Networks with higher $T_p$ and $T_C$, such as Pegasus and Zephyr, are particularly advantageous for QA, since device noise can hinder exploration of the spin-glass regime at very low temperatures. 

Understanding how topology shapes the energy landscape is essential from both technical and thermodynamic perspectives. From a technical standpoint, it informs the design of qubit architectures that can efficiently embed complex optimization problems. From a thermodynamic viewpoint, it reveals how bottlenecks and barrier distributions influence relaxation and equilibration, providing measurable benchmarks for QA performance.

Rugged landscapes with multiple metastable states serve as natural testbeds for assessing the performance of quantum annealers and for identifying conditions under which a genuine quantum speedup may emerge. When optimization problems are mapped onto spin-glass Hamiltonians, the structure of energy barriers and frustration patterns directly encodes the computational hardness of the instance. Examining percolation, cluster formation, and spin dynamics across different network topologies, therefore, provides a unified framework for designing architectures that are both computationally efficient and capable of generating energy landscapes conducive to exploring quantum advantages.


\section*{Acknowledgments}
We thank Exacore HPC Uniandes for providing high performance computing time. GT acknowledges support from Fondo de Investigaciones, Facultad de Ciencias, Universidad de los Andes INV-2023-176-2951. FJGR acknowledges financial support from the Spanish Government via the project PID2024-161371NB-C21 (MCIU/AEI/FEDER, EU).
\bibliography{references.bib}
\end{document}